\documentclass[twocolumn,preprintnumbers]{revtex4}
\usepackage{epsf}
\usepackage{amsmath,amssymb}
\newcommand{\bvec}{\boldsymbol}

\begin{document}
\preprint{KUNS-2684}
\title{Tetrahedral $4\alpha$ and $^{12}\textrm{C}+\alpha$ cluster structures in $^{16}$O}
\author{Yoshiko Kanada-En'yo}
\affiliation{Department of Physics, Kyoto University, Kyoto 606-8502, Japan}
\begin{abstract}
We have investigated structures of the ground and excited states of $^{16}\textrm{O}$
with the method of variation after spin-parity projection in the antisymmetrized molecular dynamics model
combined with the generator coordinate method of $^{12}\textrm{C}+\alpha$ cluster.
The calculation reasonably reproduces the experimental
energy spectra,  $E2$, $E3$, $E4$, $IS1$ transitions, and $\alpha$-decay properties.
The formation of 4 $\alpha$ clusters has been confirmed from nucleon degrees of freedom in the AMD model
without assuming existence of any clusters.
They form ``tetrahedral'' $4\alpha$- and 
 $^{12}\textrm{C}+\alpha$-cluster structures.
The $^{12}\textrm{C}+\alpha$ structure constructs 
the $K^\pi=0^+$ band consisting of the $0^+_2$, $2^+_1$, and $4^+_1$ states
and the $K^\pi=0^-$ band of the $1^-_2$, $3^-_2$, and $5^-_1$ states. 
The $0^+_1$, $3^-_1$, and $4^+_2$ states are assigned to the ground band constructed from 
the tetrahedral 4$\alpha$ structure. The $0^+_1$ and $3^-_1$ are approximately interpreted as 
$T_d$ band members with the ideal tetrahedral configuration. 
The ground state $4\alpha$ correlation plays an important role in  enhancement of the $E3$ transition strength 
to the $3^-_1$. The $4^+_2$ state is not the ideal $T_d$ member but constructed from 
a distorted tetrahedral $4\alpha$ structure. Moreover, 
significant state mixing of the tetrahedral $4\alpha$ and  $^{12}\textrm{C}+\alpha$ cluster structures 
occurs between $4^+_1$ and $4^+_2$ states, indicating that the $T_d$ configuration of $4\alpha$ is 
rather fragile at $J^\pi=4^+$.
\end{abstract}


\maketitle

\section{Introduction}\label{sec:introduction}

Nuclear deformation is one of typical
collective motions in nuclear systems.
Not only axial symmetric quadrupole deformations 
but also triaxial and octupole deformations have been attracting interests.
In light nuclear systems, further exotic shapes have been expected because of cluster structures. 
For instance, a triangle shape in $^{12}$C and a tetrahedral one in $^{16}$O have been discussed 
by assuming $3\alpha$- and $4\alpha$-cluster structures.
In old days, non-microscopic $\alpha$-cluster models have been applied in order 
to understand energy spectra of $^{12}$C and $^{16}$O \cite{wheeler37,dennison54}. 
Wheeler has suggested a low-lying $3^-$ state of $^{16}$O
as vibration of the tetrahedral configuration of 4 $\alpha$ particles
\cite{wheeler37}. This state has been assigned to 
the lowest negative-parity state $^{16}$O($3^-_1$, 6.13 MeV), which has been experimentally established later. 
Since 1970's, microscopic and semi-microscopic 
cluster models  have been applied in order to investigate 
cluster structures of $^{16}$O \cite{brink70,Suzuki:1976zz,Suzuki:1976zz2,fujiwara80,Libert-Heinemann:1980ktg,bauhoff84,Descouvemont:1987uvu,Descouvemont:1991zz,Descouvemont:1993zza,Fukatsu92,Funaki:2008gb,Funaki:2010px,Yamada:2011ri,Kanada-En'yo:2013dma,Horiuchi:2014yua}.

The $T_d$ symmetry of the tetrahedral  $4\alpha$-cluster structure 
has been discussed for a long time to understand energy spectra of  $^{16}$O
\cite{wheeler37,brink70,bauhoff84,Robson:1979zz,Elliott:1985dfe}. 
The ideal tetrahedral 4$\alpha$ configuration with the $T_d$ symmetry
constructs a rotational band of $0^+$, $3^-$, $4^+$, $\ldots$, states.
The ground state and the $3^-_1$ at 6.13 MeV have been assigned to the 
$T_d$-invariant 4$\alpha$ band. This assignment is supported 
by the observed strong $E3$ transition ~\cite{Robson:1979zz,Buti:1986zz} 
and $\alpha$-transfer cross sections on $^{12}$C~\cite{Elliott:1985dfe}. 
However, the assignment of the $4^+$ state in the tetrahedral 4$\alpha$  band has not 
been confirmed yet. Robson assigned the $4^-_1$ at 10.36 MeV  as the $T_d$ band
\cite{Robson:1979zz}. This assignment describes the significantly strong $E4$ transition.
However, it contradicts to 
the large $\alpha$-decay width of the $4^-_1$. Alternatively, 
the $4^-_1$(10.36 MeV) has been considered to belong to a $^{12}{\rm C}+\alpha$ cluster band starting from 
the $0^+_2$(6.05 MeV)
\cite{Suzuki:1976zz,Suzuki:1976zz2,Libert-Heinemann:1980ktg,Descouvemont:1987uvu,Descouvemont:1991zz,Descouvemont:1993zza}.
The strong $\alpha$-transfer and weak 
two-nucleon transfer cross sections for the $4^+_1$(10.36 MeV) support 
the $^{12}{\rm C}+\alpha$ 
cluster structure with the predominant $4p4h$ component \cite{Zisman:1970ey,Lowe:1972ked,Becchetti:1980nwj}. 
Elliott has discussed $\alpha$-transfer cross sections 
and assigned the $0^+_1$, $3^-_1$(6.13 MeV), and $4^+_2$(11.10 MeV) to the $T_d$ band constructed from 
a tetrahedrally deformed intrinsic state \cite{Elliott:1985dfe}.
Very recently, algebraic approaches for the $4\alpha$ system has been revived by Bijker and Iachello 
\cite{Bijker:2014tka,Bijker:2016bpb}
to describe the experimental energy spectra of $^{16}$O based on the $T_d$ symmetry and its excitation modes, 
which has been proposed by
Wheeler. In their works, the $4^+_1$(10.36 MeV) was assigned again to the 
$T_d$ band. Although the $4\alpha$ models in Refs.~\cite{Robson:1979zz,Bijker:2014tka,Bijker:2016bpb} describe 
the experimental $B(E4)$ for $0^+_1 \to 4^+_1$(10.36 MeV), the calculated form factors disagree to the experimental 
data measured by $(e,e')$ scattering as already pointed out in Ref.~\cite{Buti:1986zz}. 

In addition to the tetrahedral $4\alpha$ structure, 
$^{12}{\rm C}+\alpha$ cluster states appear in a similar energy region. 
The lowest positive-parity $^{12}{\rm C}+\alpha$ band is the $K^\pi=0^+$ band and its counterpart of
the parity doublet is the $K^\pi=0^-$ band \cite{horiuchi68}.
The $0^+_2$(6.06 MeV), $2^+_1$ (6.92 MeV), $4^+_1$(10.36 MeV), and $6^+_2$(16.23 MeV)
are assigned to the $K^\pi=0^+$ band, and the $1^-_2$(9.59 MeV), $3^-_2$(11.60 MeV), 
and $5^-_1$(14.66 MeV) are assigned to the $K^\pi=0^-$ band
because these states have similar features of $\alpha$-transfer and 
$\alpha$-decay properties.

In spite of rich cluster phenomena in $^{16}$O, there is no microscopic calculation
that sufficiently describes low-energy spectra and cluster structures of $^{16}$O.
Firstly, most of microscopic cluster model calculations fail to reproduce 
excitation energies of the $^{12}\textrm{C}+\alpha$ cluster states except for the case using 
a particularly strong exchange nuclear interaction \cite{bauhoff84}.
For instance, in $^{12}\textrm{C}+\alpha$ and $4\alpha$ cluster model calculations,  
the band-head energy of the  $K^\pi=0^+$ $^{12}\textrm{C}+\alpha$ 
band is calculated to be about $E_x(0^+_2)=16$ MeV, 
which largely overestimates the experimental value $E_x(0^+_2)=6.06$ MeV \cite{fujiwara80,Descouvemont:1993zza}. 
Therefore, it was difficult to solve the problem of possible coexistence of  
$^{12}\textrm{C}+\alpha$ and  tetrahedral $4\alpha$ states in a similar energy region.  
Secondly, both non-microscopic and microscopic cluster models {\it a priori} 
assume existence of $\alpha$ clusters and/or a $^{12}$C cluster, and are not able to check whether clusters are 
actually formed from nucleon degrees of freedom.
In mean-field calculations, 
the spherical $p$-shell closed state is usually obtained for the ground state solution
except for the cases using particularly strong exchange nuclear interactions \cite{eichler70,onishi71,takami95}. 
Even though significant mixing of 
higher-shell components in the ground state of $^{16}$O has 
been suggested in recent works with extended mean-field and shell-model approaches 
\cite{Bender:2002hba,Dytrych:2007sv,Utsuno:2011zz},
neither intrinsic shape nor cluster structure has been discussed explicitly. Moreover,
it is generally difficult for mean-field approaches to describe well developed cluster structures in excited states.

Recently, we applied the antisymmetrized molecular dynamics (AMD) method 
\cite{KanadaEnyo:1995tb,KanadaEnyo:1995ir,AMDsupp,KanadaEn'yo:2012bj}
to $^{16}$O and 
found a tetrahedral shape with the 4$\alpha$-cluster structure in a fully microscopic calculation 
based on nucleon degrees of freedom 
without assuming existence of clusters \cite{KanadaEn'yo:2012nw}. 
More recently,  in a first principle 
calculation using the chiral nuclear
effective field theory,  the tetrahedral $4\alpha$ structure has been found 
in the ground state of  $^{16}$O\cite{Epelbaum:2013paa}.

Our aim is to investigate cluster structures of low-lying states of $^{16}$O. 
We focus on the possible coexistence of the 
tetrahedral 4$\alpha$ and $^{12}{\rm C}+\alpha$ states and discuss, in particular, 
$4^+$ states in the tetrahedral 4$\alpha$ and $^{12}{\rm C}+\alpha$ bands.
To answer to the questions whether 16 nucleons form 4 $\alpha$ clusters and whether they 
are arranged in a tetrahedral configuration, we first apply the method of variation after spin-parity projection (VAP) 
in the framework of AMD, which we call the AMD+VAP
\cite{Kanada-Enyo:1998onp}. 
Then we combine the AMD+VAP
with the generator coordinate method (GCM) \cite{GCM} of $^{12}{\rm C(AMD)}+\alpha$, in which 
we use the $^{12}{\rm C(AMD)}$ cluster wave functions obtained by the AMD+VAP for 
$^{12}$C. 
The AMD+VAP method has been proved to be useful to describe
structures of light nuclei and succeeded to reproduce properties of the ground and excited states 
of $^{12}$C~\cite{Kanada-Enyo:1998onp,KanadaEn'yo:2006ze}. 
The $^{12}{\rm C(AMD)}+\alpha$ GCM has been applied in our previous work 
to investigate positive-parity states of $^{16}$O \cite{Kanada-En'yo:2013dma}. 
By combining the AMD+VAP with the $^{12}{\rm C(AMD)}+\alpha$ GCM, we obtain 
better description of asymptotic behavior
and excitation energies of $^{12}{\rm C}+\alpha$ states. 
We calculate the positive- and negative-parity levels, transition strengths, and 
$\alpha$-decay widths, and discuss cluster structures of $^{16}$O.

The paper is organized as follows. In the next section, the framework of the 
present calculation is explained. Section~\ref{sec:interaction} describes
the adopted effective interactions. 
The results of $^{16}$O are shown in Sec.~\ref{sec:results}.
Finally, we give a summary and outlooks in Sec.~\ref{sec:summary}.

\section{Formulation} \label{sec:formulation}
The present method is the AMD+VAP combined with the 
$^{12}{\rm C(AMD)}+\alpha$ GCM. The details of 
the AMD+VAP are described in Refs.~\cite{Kanada-Enyo:1998onp,KanadaEn'yo:2006ze}. 
For the formulation of the $^{12}{\rm C(AMD)}+\alpha$ GCM, 
the reader is referred to Ref.~\cite{Kanada-En'yo:2013dma}.

\subsection{AMD wave function and VAP}

We define the AMD model wave function and perform 
energy variation to obtain the energy-minimum solution in the AMD model space.
An AMD wave function for an $A$-nucleon system 
is given by a Slater determinant of Gaussian wave packets,
\begin{equation}
 \Phi_{\rm AMD}(\bvec{Z}) = \frac{1}{\sqrt{A!}} {\cal{A}} \{
  \varphi_1,\varphi_2,...,\varphi_A \},
\end{equation}
where the $i$th single-particle wave function is written by a product of the
spatial, intrinsic spin, and isospin wave functions as
\begin{align}
 \varphi_i&= \phi_{\bvec{X}_i} \chi^\sigma_i \chi^\tau_i,\\
 \phi_{\bvec{X}_i}({\bvec{r}}_j) & =   \left(\frac{2\nu}{\pi}\right)^{4/3}
\exp\bigl\{-\nu({\bvec{r}}_j-\frac{\bvec{X}_i}{\sqrt{\nu}})^2\bigr\},
\label{eq:spatial}\\
 \chi^\sigma_i &= (\frac{1}{2}+\xi_i)\chi_{\uparrow}
 + (\frac{1}{2}-\xi_i)\chi_{\downarrow}.
\end{align}
Here, $\phi_{\bvec{X}_i}$ and $\chi^\sigma_i$ are the spatial and intrinsic spin functions, and 
$\chi^\tau_i$ is the isospin
function fixed to be proton or neutron. The width parameter $\nu$ 
is chosen to be a common value.

Thus, the AMD wave function
is specified by a set of variational parameters, $\bvec{Z}\equiv 
\{\bvec{X}_1,\bvec{X}_2,\ldots, \bvec{X}_A,\xi_1,\xi_2,\ldots,\xi_A \}$
for Gaussian center positions ($\bvec{X}_1,\ldots, \bvec{X}_A$) and intrinsic spin orientations 
($\xi_1,\ldots,\xi_A$) of all single-nucleon 
wave functions, which are independently 
treated as variational parameters. 
In the AMD framework,  existence of neither clusters nor a core nucleus is assumed, but
nuclear structures are described based on  nucleon degrees of freedom.
Nevertheless, the AMD model space covers various cluster structures 
as well as shell-model structures owing to
flexibility of spatial configurations of single-nucleon Gaussian wave functions, 
which are fully antisymmetrized. Therefore, if a cluster structure is favored in a system, 
the cluster structure is 
automatically obtained in the energy variation. 

To express a $J^\pi$ state, an AMD wave function is projected onto the spin-parity eigenstate, 
\begin{equation}
\Phi^{J\pi}(\bvec{Z})=P^{J\pi}_{MK}\Phi_{\rm AMD}(\bvec{Z}),
\end{equation}
where $P^{J\pi}_{MK}$ is the spin-parity projection operator. 
To obtain the wave function for the $J^\pi$ state, the VAP
is performed for the $J^\pi$-projected AMD wave function,  
\begin{equation}
\delta\frac{\langle \Phi^{J\pi}(\bvec{Z})|H|\Phi^{J\pi}(\bvec{Z}) \rangle}{\langle \Phi^{J\pi}(\bvec{Z})|\Phi^{J\pi}(\bvec{Z}) \rangle}=0,
\end{equation}
with respect to variation $\delta\bvec{Z}$. 
After the VAP, we obtain the optimum parameter set $\bvec{Z}^{\rm opt}_{J^\pi}$ for the $J^\pi$ state. 
This method is called the AMD+VAP. 
The obtained AMD wave function $\Phi_{\rm AMD}(\bvec{Z}^{\rm opt}_{J^\pi})$
expressed by a single Slater determinant is regarded as the intrinsic state of the $J^\pi$ state. 
Note that the $J^\pi$-projected AMD wave function is no longer a Slater determinant
and, in principle,  contains higher correlations beyond the Hartree-Fock approach.

When a local minimum solution is obtained by the VAP for $J^\pi$, it is regarded as
the second (or higher) $J^\pi$ state. Another way to obtain 
the AMD configuration $\bvec{Z}^{\rm opt}_{J^\pi_2}$ optimized
for the $J^\pi_2$ state is 
VAP for the component orthogonal to the obtained $J^\pi_1$ state,
\begin{equation}
\Phi^{J\pi}_{\rm exc}(\bvec{Z})=\left (1- \frac{|\Phi^{J\pi}(\bvec{Z}^{\rm opt}_{J^\pi_1}) \rangle
\langle\Phi^{J\pi}(\bvec{Z}^{\rm opt}_{J^\pi_1})| }{
\langle\Phi^{J\pi}(\bvec{Z}^{\rm opt}_{J^\pi_1})|\Phi^{J\pi}(\bvec{Z}^{\rm opt}_{J^\pi_1}) \rangle}
\right) \Phi^{J\pi}(\bvec{Z}).
\end{equation}

In the AMD+VAP method, all the AMD wave functions obtained by VAP for various $J'^{\pi'}_{n'}$ 
are superposed to obtain the final wave function for the $J^\pi_n$ state, 
\begin{equation}
\Psi^{^{16}\textrm{O}(J^\pi_n)}_\textrm{VAP}
=\sum_{K}\sum_{\beta=J'^{\pi'}_{n'}} c_\textrm{vap}(J^\pi_n;K,\beta)
P^{J\pi}_{MK}\Phi_\textrm{AMD}(\bvec{Z}^{\rm opt}_\beta),
\end{equation}
where coefficients $c_\textrm{vap}(J^\pi_n;K,\beta)$ for the $J^\pi_n$ state are determined by 
diagonalization of  the norm and Hamiltonian matrices.

\subsection{$^{12}{\rm C(AMD)}+\alpha$ GCM}

In the $^{12}{\rm C(AMD)}+\alpha$ GCM, 
the $^{12}$C-$\alpha$ distance is treated as 
a generator coordinate. For the description of the  $^{12}$C cluster, 
we use $^{12}$C wave functions $\Phi^{^{12}\textrm{C}}_\textrm{AMD}(\bvec{Z}^\textrm{opt}_{\beta_\textrm{C}})$
 obtained by the AMD+VAP for $^{12}$C.
 Here the label
$\beta_\textrm{C}=J^\pi_n$ is used for the $^{12}\textrm{C}(J^\pi_n)$ state. In the present calculation 
we use three configurations, $\beta_\textrm{C}=0^+_1$, $0^+_2$, and $1^-_1$, corresponding to 
$^{12}\textrm{C}(0^+_1)$, $^{12}\textrm{C}(0^+_2)$, and  
$^{12}\textrm{C}(1^-_1)$, respectively. 
These three configurations describe well 
energy spectra of $^{12}$C as shown in Ref.~\cite{Kanada-En'yo:2013dma}.

To describe inter-cluster motion between $^{12}$C and $\alpha$ clusters, 
we superpose 
the $^{12}\textrm{C}+\alpha$ wave functions with various distance $d$ 
using the $^{12}$C-cluster wave function $\Phi^{^{12}\textrm{C}}_\textrm{AMD}
(-{\bvec{d}}/4;\bvec{Z}^\textrm{opt}_{\beta_\textrm{C}})$
 localized at a mean center-of-mass position $-{\bvec{d}}/4$ (${\bvec{d}}=(0,0,d)$)
and 
the $(0s)^4$ $\alpha$-cluster wave function $\Phi_\alpha(3{\bvec{d}}/4)$
at $3{\bvec{d}}/4$.
A $^{12}$C-cluster configuration, $\Phi^{^{12}\textrm{C}}_\textrm{AMD}(\bvec{Z}^\textrm{opt}_{\beta_\textrm{C}})$, has
a cluster structure with an intrinsic deformation oriented in a specific direction. 
To take into account angular momentum projection of the subsystem $^{12}$C, we consider 
rotation $\tilde{R}(\Omega)$ of the 
$\Phi^{^{12}\textrm{C}}_\textrm{AMD}
(-{\bvec{d}}/4;\bvec{Z}^\textrm{opt}_{\beta_\textrm{C}})$ around $-{\bvec{d}}/4$.

The total wave function for $^{16}\textrm{O}(J^\pi_n)$ of the $^{12}$C(AMD)+$\alpha$ GCM model
 is written as
\begin{eqnarray}
&&\Psi_{\alpha{\rm GCM}}^{J^\pi_n}\nonumber\\
&&=\sum_{K,i,j,\beta_\textrm{C}}c_\textrm{gcm}(J^\pi_n;K,i,j,\beta_\textrm{C})
\Phi^{J\pi K}_{^{12}{\rm C}+\alpha}(d_i,\Omega_j,\bvec{Z}^\textrm{opt}_{\beta_\textrm{C}}),\nonumber \\
&&\\
&&\Phi^{J\pi K}_{^{12}{\rm C}+\alpha}(d,\Omega_j,\bvec{Z}^\textrm{opt}_{\beta_\textrm{C}})\nonumber \\
&&\equiv P^{J\pi}_{MK}{\cal A}\left\{\tilde{R}(\Omega)
\Phi_{^{12}{\rm C}}^{\rm AMD}(-{\bvec{S}}/4;\bvec{Z}^\textrm{opt}_{\beta_\textrm{C}})\cdot \Phi_\alpha(3{\bvec {d}}/4) \right\},\nonumber\\
\end{eqnarray}
where coefficients $c_\textrm{gcm}(J^\pi_n;K,i,j,\beta_\textrm{C})$ are determined by 
solving the Hill-Wheeler equation \cite{GCM}, i.e. diagonalizing the norm and Hamiltonian matrices.
The superposition of rotated $^{12}$C-cluster wave functions is equivalent to 
linear combination of various spin states of the $^{12}$C cluster projected from the intrinsic state. 
In addition to 
 $^{12}$C-cluster rotation,
excitation of the $^{12}$C cluster is taken into account
by superposing configurations, $\beta_\textrm{C}=0^+_1$, $0^+_2$, $1^-_1$. 
Moreover, $3\alpha$ breaking is 
already taken into account in 
$\Phi^{^{12}\textrm{C}}_\textrm{AMD}(\bvec{Z}^\textrm{opt}_{\beta_\textrm{C}})$.

\subsection{AMD+VAP combined with $^{12}\textrm{C(AMD)}+\alpha$ GCM}
We combine the AMD+VAP method with the $^{12}\textrm{C(AMD)}+\alpha$ GCM
by superposing all basis wave functions,  
\begin{eqnarray}
&&\Psi^{^{16}\textrm{O}(J^\pi_n)}_{\textrm{VAP}+\alpha\textrm{GCM}}\nonumber\\
&&=\sum_{K,\beta} c_\textrm{vap}(J^\pi_n;K,\beta)
P^{J\pi}_{MK}\Phi_\textrm{AMD}(\bvec{Z}^{\rm opt}_\beta)\\
&&+\sum_{K,i,j,\beta_\textrm{C}}c_\textrm{gcm}(J^\pi_n;K,i,j,\beta_\textrm{C})
\Phi^{J\pi K}_{^{12}{\rm C}+\alpha}(d_i,\Omega_j,\bvec{Z}^\textrm{opt}_{\beta_\textrm{C}}),\nonumber\\
\end{eqnarray}
where coefficients, $c_\textrm{vap}$ and $c_\textrm{gcm}$, are determined by the diagonalization of the norm and Hamiltonian matrices.
We call this method ``VAP+$\alpha$GCM,''.

\section{Effective nuclear interactions} \label{sec:interaction}
In the present calculation, we use
the effective nuclear interactions with the parametrization same
as that used for $^{12}$C
in the AMD+VAP calculation \cite{Kanada-Enyo:1998onp,KanadaEn'yo:2006ze}. 
They are the MV1 central force \cite{MVOLKOV} and the G3RS spin-orbit force \cite{LS1,LS2}.
The MV1 force contains finite-range two-body and zero-range three-body terms.
We use the case-1 parametrization of the MV1 force and set 
the Bartlett ($b$), Heisenberg ($h$), and Majorana ($m$) parameters as $b=h=0$ and $m=0.62$.
As for strengths of the two-range Gaussian of the G3RS spin-orbit force, we use
$u_{I}=-u_{II}\equiv u_{ls}=3000$ MeV to reproduce the $2^+_1$ excitation energy of $^{12}$C
with the MV1 force.
The Coulomb force is approximated using a seven-range
Gaussian form. 

With these interactions, 
properties of the ground and excited states 
of $^{12}$C are described well by the AMD+VAP calculation \cite{Kanada-Enyo:1998onp,KanadaEn'yo:2006ze}.
As for a symmetric nuclear matter,  the MV1 force with the present parametrization
gives the saturation density 
$\rho_s=0.192$ fm$^{-3}$, the saturation energy $E_s=-17.9$ MeV, 
the effective nucleon mass $m^*_\textrm{SNM}=0.59m$, and  
the incompressibility $K=245$ MeV. 

It is known that usual two-body effective nuclear interactions with 
mass-independent parameters have an
overshooting problem of nuclear binding and density with increase of the mass number
and are not able to describe the saturation property. 
The overshooting problem is improved with the use of the MV1 force, because
it contains a zero-range three-body force, which is equivalent 
to a density-dependent force for spin and isospin saturated systems.
In the sense that the MV1 force consists of finite-range two-body and ``density-dependent'' zero-range forces,  
it can be categorized to a similar type interaction to Gogny forces.

The present interaction parameters gives reasonable result for 
the $\alpha$, $^{12}$C, and $^{16}$O bindings 
compared with the experimental binding energies (B.E.)
of $\alpha$ (28.30 MeV), $^{12}$C (92.16 MeV), and $^{16}$O (127.62 MeV):
the calculated B.E. of the  $(0s)^4$ $\alpha$ particle is  $27.8$ MeV, that of 
$^{12}$C obtained by the AMD+VAP with 3 configurations 
($^{12}{\rm C}(0^+_1)$, $^{12}{\rm C}(0^+_2)$, and $^{12}{\rm C}(1^-_1)$) 
is $87.6$ MeV, and that of $^{16}$O with the AMD+VAP (VAP+$\alpha$GCM) is 123.0 (123.5) MeV. 
The calculated $\alpha$-decay threshold of $^{16}$O is about 8 MeV, 
which is in reasonable agreement with the experimental value 7.16 MeV.

\section{Results}\label{sec:results}
\subsection{Procedure and parameter setting}

The width parameter $\nu$ for all wave functions of  $\alpha$,  $^{12}$C, and  
$^{16}$O is chosen to be a common value so that 
the center of mass motion can be exactly removed.  In the present calculation, we use 
$\nu=0.19$ fm$^{-2}$, which minimizes 
the energies of $^{12}$C and $^{16}$O.

In the AMD+VAP calculation of $^{16}$O, we obtain 
9 configurations ($\beta$) for $J^\pi_n=0^+_{1,2}$, 
$2^+_1$, $4^+_{1,2}$, $1^-_{1}$, $2^-_{1}$, $3^-_{1}$, and 
$5^-_1$. First we obtain the $0^+_{1}$ configuration $\bvec{Z}^{\rm opt}_{0^+_1}$
with the VAP without the orthogonal condition, and next
obtain the $0^+_{2}$ configuration $\bvec{Z}^{\rm opt}_{0^+_2}$
with the VAP with the condition orthogonal to the $0^+_{1}$.
For other $J^\pi$, we iteratively achieve the VAP without the orthogonal condition 
by starting from $\bvec{Z}^{\rm opt}_{0^+_1}$ and $\bvec{Z}^{\rm opt}_{0^+_2}$ 
as initial configurations. 
In the VAP for $J^\pi=4^+$, we found minimum and local minimum solutions for
the $4^+_{1}$ and $4^+_2$ configurations. For $J^\pi$ other than $4^+$, 
we did not obtain local minimum solutions but obtained only a minimum solution in two cases of 
initial configurations.

In the $^{12}\textrm{C(AMD)}+\alpha$ GCM,  
we use inter-cluster distances $d_i=1.2,\ 2.4,\ 3.6, \ldots,8.4$ fm (7 points with 1.2 fm interval) 
for $\beta_\textrm{C}=0^+_1$ and $0^+_2$ of the $^{12}\textrm{C}$ configurations. 
For $\beta_\textrm{C}=1^-_1$, we adopt 
$d_i=1.2,2.4,3.6\cdots,6.0$ fm (5 points with 1.2 fm interval) 
to save computational costs. 
For Euler angles $\Omega_j$ of the $^{12}$C-cluster rotation $\tilde{R}(\Omega_j)$ we use seventeen points
$(j=1,\ldots,17)$, as described in Ref.~\cite{Kanada-En'yo:2013dma}.

In the $K$-mixing, we truncate $|K| \ge 5 $ components to save computational resources.

As described previously, we combine the AMD+VAP and  $^{12}{\rm C(AMD)}+\alpha$ GCM
to obtain final result. 
In the AMD+VAP, 
each $^{16}$O wave function is essentially expressed by the 
$J^\pi$ state projected from a Slater determinant, and therefore, 
it is useful to discuss an intrinsic shape of the state. 
In other words, strong-coupling cluster structures are 
obtained within the AMD+VAP. 
On the other hand, the $^{12}{\rm C(AMD)}+\alpha$ GCM 
is essential to describe weak-coupling $^{12}{\rm C}+\alpha$ cluster states, 
for which the angular momentum 
projection of the subsystem $^{12}{\rm C}$ is necessary.
In the present paper, we start from the AMD+VAP result (hereafter we call it the VAP result)
and then analyze the VAP+$\alpha$GCM result to discuss how the 
VAP result is affected by mixing of 
$^{12}{\rm C(AMD)}+\alpha$ configurations.
Note that the obtained VAP states show predominantly $4\alpha$ structures, which are 
approximately included by the $^{12}{\rm C(AMD)}+\alpha$ model space.

\subsection{Energies, radius, and transitions}

The calculated and experimental values for 
B.E., root-mean-square(r.m.s.) radius, and $^{12}\textrm{C}+\alpha$ threshold  
are listed in Table \ref{tab:energy}. 
The ground state properties calculated by the VAP and VAP+$\alpha$GCM
are similar to each other, and they are in reasonable agreement with the experimental data.

\begin{table}[ht]
\caption{The calculated and experimental values of B.E., r.m.s. radius, and the $^{12}\textrm{C}+\alpha$ 
threshold energy. The experimental r.m.s radius is reduced from the charge radius
in Ref.~\cite{angeli13}
\label{tab:energy}
}
\begin{center}
\begin{tabular}{cccc}
\hline
 & VAP & VAP+$\alpha$GCM & exp. \\
B.E. (MeV)	&	123.0 & 123.5 &	127.62 \\
r.m.s. radius (fm) & 2.69  & 2.73 & 2.55 \\
$^{12}\textrm{C}+\alpha$ (MeV) & 7.6  & 8.2 & 7.16 \\
\hline
\end{tabular}
\end{center}
\end{table}

\begin{figure}[th]
\epsfxsize=0.4\textwidth
\centerline{\epsffile{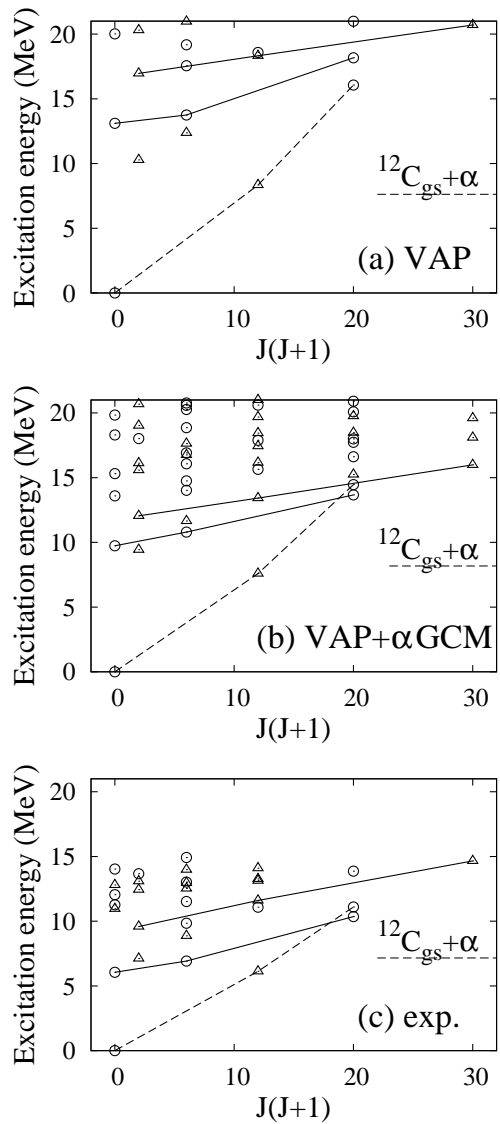}}
\caption{\label{fig:o16rot}
Energy levels of $^{16}$O calculated by (a) the VAP, (b) VAP+$\alpha$GCM, 
and (c) those of the experimental data. 
The positive- and negative-parity levels are shown by circles and triangles, respectively. 
The ground and $^{12}\textrm{C}+\alpha$  bands are 
connected by dashed and solid lines, respectively.  
}
\end{figure}

Energy levels are shown in Fig.~\ref{fig:o16rot}.
In the figure, 
the energy levels in the ground and $^{12}\textrm{C}+\alpha$ bands are connected by dashed and solid lines,
respectively.
In a usual assignment, the experimental $0^+_2$, $2^+_1$, $4^+_1$, $1^-_2$, $3^-_2$, and $5^-_1$ states
are considered to belong to the 
$K^\pi=0^\pm$ $^{12}\textrm{C}+\alpha$ cluster bands from 
$\alpha$-decay and $\alpha$-transfer properties of these states.
For the ground band, 
we tentatively assign the experimental $0^+_1$, $3^-_1$, and $4^+_2$ states as band members following the 
assignment of Ref.~\cite{Elliott:1985dfe}. 
In the VAP and VAP+$\alpha$GCM results, we can 
categorize calculated energy levels 
into the ground and $^{12}\textrm{C}+\alpha$ bands 
based on  $E2$ transition properties as well as analysis of intrinsic structures. 
In the VAP calculation, the $4^+$, $1^-$, and $3^-$ states of the $^{12}\textrm{C}+\alpha$ band 
are not obtained as local minimum solutions, but they are constructed by the $J^\pi$ projection from the
$^{12}\textrm{C}+\alpha$ cluster structure obtained for $0^+_2$, $2^+_1$, and $5^-_1$. 

Excitation energies of the $^{12}\textrm{C}+\alpha$ states are much overestimated by 
the VAP calculation compared with the experimental data.
The $K^\pi=0^+$ band-head energy of the VAP is $E_x(0^+_2)=13.1$ MeV, which is about twice higher than the experimental value (7.16 MeV).  
The VAP+$\alpha$GCM calculation gives a better result for 
the $^{12}\textrm{C}+\alpha$ band energies owing to rotation and internal excitation of the 
$^{12}\textrm{C}$ cluster.
The VAP+$\alpha$GCM gives $E_x(0^+_2)=9.7$ MeV of the band-head energy. 
The energy 
is still higher than the experimental value, but the overestimation 
is significantly improved by the present VAP+$\alpha$GCM calculation compared with the 
theoretical value $E_x(0^+_2)\sim 16$ MeV of 
microscopic cluster model calculations with the Volkov interaction.
As a result of the significant energy reduction of the $^{12}\textrm{C}+\alpha$ states, the ordering of the ground 
and $^{12}\textrm{C}+\alpha$ bands is reversed at $J^\pi=4^+$ from the VAP to the VAP+$\alpha$GCM.
The $4^+_1$ state is the ground band member in the VAP, whereas it belongs to the $K^\pi=0^+$ 
$^{12}\textrm{C}+\alpha$ band in the VAP+$\alpha$GCM consistently to the usual assignment of the experimental levels
and also that of Ref.~\cite{Elliott:1985dfe}.
Strictly speaking, state mixing occurs between the $4^+_1$ and $4^+_2$ states as discussed later.

In Table \ref{tab:be2}, 
the calculated $E2$, $E3$, $E4$, and isoscalar dipole ($IS1$) transition strengths are 
shown compared with experimental data. We also show the theoretical values of 
Ref.~\cite{Suzuki:1976zz2} calculated by 
a semi-microscopic $^{12}\textrm{C}+\alpha$ cluster model with the 
orthogonal condition model (OCM) \cite{OCM}.
In the VAP result, the $E2$ transition for $4^+_1\to 2^+_1$ is weak because these states belong to
different bands, which contradicts to the experimental strong $E2$ transition. 
On the other hand,  the $B(E2)$ value calculated by the  
VAP+$\alpha$GCM is as large as the experimental data. Relatively large $B(E2)$ values for the
in-band transitions in the $^{12}\textrm{C}+\alpha$ bands are consistent with experimental data.

The $E3$ transition strength for $3^-_1\to 0^+_1$ is considerably large in both calculations
because of the dominant tetrahedral 4$\alpha$ component in the $0^+_1$ and $3^-_1$ states. 
The calculated $B(E3)$ is in good agreement with the experimental data. 
It should be pointed out that 
the ground state $4\alpha$ correlation gives important contribution to the enhancement of $B(E3;3^-_1\to 0^+_1)$.
Indeed, if we assume the $p$-shell closed configuration of the final $0^+$ state, 
the strength from the $3^-_1$ state of the VAP+$\alpha$GCM becomes as small as $B(E3)=26$ $e^2$fm,
indicating that higher shell components contribute to the dominant part of the $E3$ strength.
The $^{12}\textrm{C}+\alpha$ OCM calculation fails to reproduce the large $B(E3)$ value because 
spatially developed $3\alpha$ configurations in the $^{12}\textrm{C}$ cluster are ignored  
in the calculation.

The calculated values of $B(E4)$ for $4^+_1\to 0^+_1$ are consistent with 
the experimental value in both of the VAP and VAP+$\alpha$GCM calculations.
Naively, it seems to contradict to the different assignments of
the $4^+_1$ state in two calculations. In the VAP result, the $4^+_1$ state belongs to the ground band with a
dominant tetrahedral $4\alpha$ component. 
The tetrahedral intrinsic state gives $B(E4)=260$ $e^2$fm$^8$ in the VAP. 
On the other hand, in the VAP+$\alpha$GCM, the $4^+_1$ is dominated by  
the $^{12}\textrm{C}+\alpha$ component different from the dominant tetrahedral 4$\alpha$ component of the $0^+_1$.
However, in the VAP+$\alpha$GCM result, 
mixing of the tetrahedral $4\alpha$ and $^{12}\textrm{C}+\alpha$ components occurs in the $4^+_1$ and 
$4^+_2$ states and enhances the $B(E4;4^+_1\to 0^+_1)$ value. 
Moreover, slight mixing of the
$^{12}\textrm{C}+\alpha$ component in the $0^+_1$ also increases the $E4$ strength in the VAP+$\alpha$GCM.
Consequently, the calculated $B(E4;4^+_1\to 0^+_1)$ is $B(E4;4^+_1\to 0^+_1)=360$  $e^2$fm$^8$ in the VAP+$\alpha$GCM.
Note that the ground state $4\alpha$ correlation gives significant contribution 
to the $E4$ transition. If we assume the $p$-shell closed $0^+$ state,  
the $B(E4)$ values for the $4^+_1$ state of the VAP is reduced to be $B(E4)=17$ $e^2$fm$^8$ 
and that of the VAP+$\alpha$GCM  is $B(E4)=4$ $e^2$fm$^8$. It indicates again that 
higher shell components enhance the $E4$ strength.
In the experimental measurement of $E4$ transitions by $(e,e')$ scattering \cite{Buti:1986zz}, it has been reported that 
the $E4$ transition strength for $0^+_1 \to 4^+_2$ is the same order as that for $0^+_1 \to 4^+_1$. 
It may support the strong state mixing between two $4^+$ states.

For the $IS1$ strength, it has been experimentally known that the low-energy $IS1$ strength exhausts
significant fraction of the energy weighted sum rule (EWSR) \cite{Harakeh:1981zz}.
The present calculation gives the significant $IS1$ strength for $0^+_1\to 1^-_1$ with $4-5$ \% of the EWSR, 
which is consistent with the experimental data.

\begin{table}[ht]
\caption{
\label{tab:be2}
$E2$, $E3$, $E4$, and $IS1$ transition strengths in $^{16}$O. 
The calculated and experimental values, and 
theoretical values of Ref.~\cite{Suzuki:1976zz2} are listed.
For the $IS1$ transition strengths, 
values in parentheses are ratio to the energy weighted sum rule \cite{Harakeh:1981zz}.  
Experimental data are taken from 
Refs.~\cite{Tilley:1993zz,Harakeh:1981zz}.}
\begin{center}
\begin{tabular}{ccccc}
\hline
$J^\pi_i\to J^\pi_f$ &VAP&VAP+$\alpha$GCM & exp. & Ref.~\cite{Suzuki:1976zz2} \\
  \multicolumn{2}{c} {$B(E2)$ (e$^2$fm$^4$)} &&\\
$2^+_1\to 0^+_1$	& 4.5 & 3.1 &	7.4$\pm$0.2	&	2.48	\\
$2^+_1\to 0^+_1$	& 51 & 141&	65$\pm$7	&	60.1	\\
$2^+_2\to 0^+_1$	& 0.1	& 2.5 &0.07$\pm$0.007	&	0.489	\\
$2^+_2\to 0^+_2$	& 8.4  & 44 &	2.9$\pm$0.7	&	4.64	\\
$4^+_1\to 2^+_1$	& 5.7 & 180 &	156$\pm$14	&	96.2	\\
$4^+_2\to 2^+_1$	& 58 & 68 &	$-$	&	$-$	\\
$1^-_1	\to 	3^-_1$ &	27 	&	35 	&	50	$\pm$	12	&	27.6	\\
$2^-_1	\to 	1^-_1$	 &	8.5 	&	10.6 	&	25 	$\pm$	4 	&	17.5	\\
$2^-_1	\to 	3^-_1$	 &	11.4 	&	14.9 	&	20 	$\pm$	2 	&	9.74	\\
$1^-_2	\to 	3^-_1$	 &	1.5 	&	0.9 	&$-$	&	$-$	\\
  \multicolumn{2}{c} {$B(E3)$ (e$^2$fm$^6$)} &&\\
$3^-_1	\to 	0^+_1$	&	191 	&	218 	&	205	$\pm$	11	&	29.6	\\
$3^-_1	\to 	0^+_2$	&	19.2 	&	0.3 	&		$-$	&	$-$	\\
 &   \multicolumn{2}{c} {$B(E4)$ (e$^2$fm$^8$)} &&\\
$4^+_1\to	0^+_1$ &	260 	&	360 	&	380	$\pm$	130	&		\\
$4^+_1\to 	0^+_2$	&	430 	&	$4.0\times 10^4$ 	&		$-$	&	$-$	\\
$4^+_2\to 	0^+_1$	&	110 	&	78 	&	$-$	&	$-$	\\
$4^+_2\to 	0^+_2$	&	$4.8 \times 10^3$ 	&	$1.50 \times 10^4$ 	&		$-$	&	$-$	\\
 &   \multicolumn{2}{c}{$B(IS1)$ (fm$^6$)}  (EWSR ratio)& & \\
$0^+_1\to 1^-_1$	&	125 (4.0\%)	&	165  (4.6\%)	&	(4.2\%)	&	$-$	\\
$0^+_1\to 1^-_2$	&	7.9	&	4.8 	&	$-$	&	$-$	\\
\hline
\end{tabular}
\end{center}
\end{table}

\subsection{Intrinsic structures}

\begin{figure}[th]
\epsfxsize=0.4\textwidth
\centerline{\epsffile{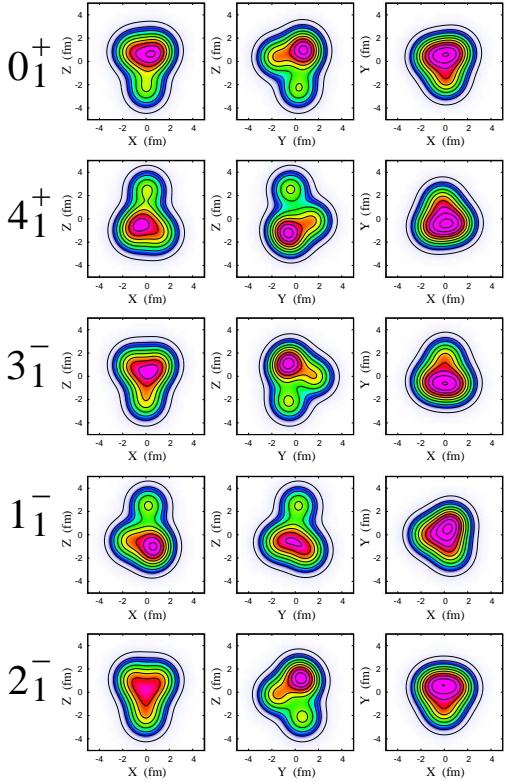}}
\caption{\label{fig:dense1}
(Color on-line) Density distributions in the intrinsic states obtained by the 
VAP for the $0^+_1$, $4^+_1$, $3^-_1$, $1^-_1$, and $2^-_1$ states.
The densities integrated along the $Y$, $X$, and $Z$ axes are plotted on the (left) $X$-$Z$, 
(middle) $Y$-$Z$, and (right) $X$-$Y$ planes, respectively. 
}
\end{figure}

\begin{figure}[th]
\epsfxsize=0.4\textwidth
\centerline{\epsffile{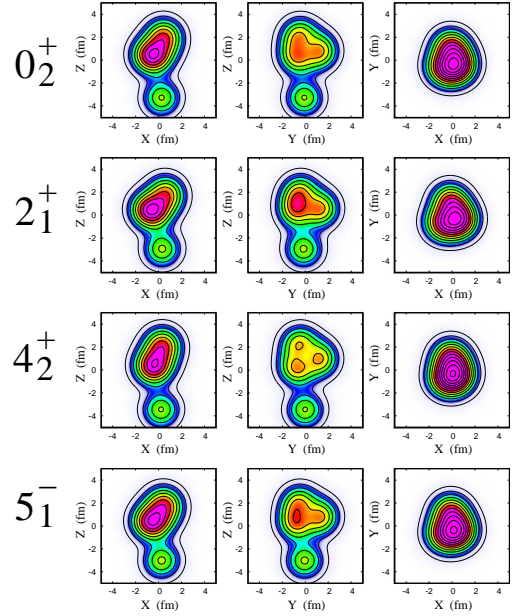}}
\caption{\label{fig:dense2}(Color on-line) Same as
Fig.~\ref{fig:dense1} but for the 
$0^+_2$, $2^+_1$, $4^+_2$, $5^-_1$ states.}
\end{figure}

Since a single AMD wave function is given by a Slater determinant,  
the AMD wave function $\Phi_\textrm{AMD}(\bvec{Z}^{\rm opt}_{J^\pi_n})$ optimized for $J^\pi_n$ in the VAP 
is regarded as the intrinsic state of the corresponding state.
Density distribution of the intrinsic states obtained by the VAP is shown 
in Figs.~\ref{fig:dense1} and \ref{fig:dense2}.

The intrinsic density shows that four $\alpha$ clusters are predominantly formed in 
the ground and excited states of $^{16}$O.
The $0^+_1$, $4^+_1$, $1^-_1$, $2^-_1$, and  $3^-_1$ states show
tetrahedral $4\alpha$ structures. The shapes are not an ideal tetrahedral configuration with the $T_d$ symmetry 
but somewhat distorted tetrahedral ones.
On the other hand, the $0^+_2$, $2^+_1$, $4^+_2$, and $5^-_1$ states
show $^{12}\textrm{C}+\alpha$ cluster structures, in which 
an $\alpha$ cluster is located far from the $^{12}$C cluster with $3\alpha$ structures.
In particular, in the  $4^+_2$ state, the $3\alpha$ structure of the $^{12}$C cluster is clearly seen and the last $\alpha$ cluster
is aligned almost on the $3\alpha$ plane. It is a similar structure to the planer $4\alpha$ configuration suggested 
in Ref.~\cite{bauhoff84} for the excited $K^\pi=0^+$ band.

In the VAP+$\alpha$GCM, the $^{16}$O wave functions are expressed by 
superposition of the VAP and $^{12}\textrm{C(AMD)}+\alpha$ wave functions. 
In general, a strong-coupling cluster having a specific intrinsic shape such as the
tetrahedral or planar $4\alpha$ structures may have a large overlap with the dominant 
configuration, whereas a weak-coupling cluster state such as the 
$^{12}\textrm{C}(0^+_1)+\alpha$ structure should contain components of various configurations.
For the $0^+_1$,  $1^-_1$,  $2^-_1$, and $3^-_1$ states, the VAP+$\alpha$GCM wave functions
contain significant components of the corresponding VAP states with more than 80\% overlap,
meaning that these states are understood as 
strong-coupling cluster states with the specific intrinsic shapes. 
For each state, the dominant VAP wave function can be approximately regarded as the intrinsic state.
On the other hand, the $^{12}\textrm{C}+\alpha$ band members of the VAP+$\alpha$GCM
contain the VAP component with less than 55\% and somewhat show  weak-coupling cluster features. 
As discussed later, the states in the $K^\pi=0^\pm$ 
$^{12}\textrm{C}+\alpha$ bands contain significant $^{12}\textrm{C}(0^+_1)+\alpha$ component 
with a rather large inter-cluster distance indicating that they are understood as weak-coupling 
$^{12}\textrm{C}(0^+_1)+\alpha$ states.
For $4^+$ states, 
the $4^+_1$ of the VAP+$\alpha$GCM has
minor (30\%)  VAP $4^+_1$ component and major (50\%) VAP $4^+_2$ component, whereas the $4^+_2$ of the VAP+$\alpha$GCM 
has 
dominant (55\%) VAP $4^+_1$ and minor (30\%) VAP $4^+_2$ components. It indicates that 
the level inversion occurs between the VAP and VAP+$\alpha$GCM calculations: 
the $4^+_1$ and $4^-_2$ states in the 
VAP+$\alpha$GCM are approximately 
assigned to the $^{12}\textrm{C}+\alpha$ and tetrahedral $4\alpha$ bands, respectively. 
However, they still contain minor VAP components by 30\% 
indicating significant state mixing between two $4^+$ states.

As discussed in 1937 by Wheeler \cite{wheeler37}, 
the ideal tetrahedral $4\alpha$ configuration with the $T_d$ symmetry and its vibration 
construct specific  band structures.
We here discuss how much components of the $T_d$ modes
are contained in the $0^+$, $3^-$, $4^+$, and $1^-$ states obtained in the present calculation.
We introduce the Brink-Bloch(BB) $4\alpha$-cluster wave function \cite{brink66} with the $T_d$ symmetry and that for 
the vibration mode and calculate overlap with the VAP and VAP+$\alpha$GCM wave functions.
The BB $4\alpha$-cluster wave function is written as
\begin{eqnarray}
&&\Phi^{4\alpha}({S}_1,\bvec {S}_2,\bvec {S}_3,\bvec {S}_4)\nonumber\\
&&={\cal A}\left\{\Phi_\alpha(\bvec {S}_1) 
\Phi_\alpha(\bvec {S}_2) \Phi_\alpha(\bvec {S}_3) \Phi_\alpha(\bvec {S}_4) \right\}.
\end{eqnarray}
For the BB $4\alpha$-cluster wave function $\Phi^{4\alpha}_{T_d}(d)$ 
with the $T_d$ symmetry, 
we choose the $4\alpha$ configuration $\bvec{S}_i$
\begin{eqnarray}
&&\bvec{S}_1=\frac{d}{\sqrt{3}}({1, 1, -1}),\\
&&\bvec{S}_2=\frac{d}{\sqrt{3}}({1, -1, 1}),\\
&&\bvec{S}_3=\frac{d}{\sqrt{3}}({-1,-1, -1}),\\
&&\bvec{S}_4=\frac{d}{\sqrt{3}}({-1, 1, 1}).
\end{eqnarray}
Here $d$ is the size parameter
of the $T_d$-invariant tetrahedral configuration of $4\alpha$.
For the wave function $\Phi^{4\alpha}_{T_d(F)}(d)$ corresponding to 
the mode ``F'', which is a vibration mode for the $1^-$ state 
on the $T_d$-invariant tetrahedral configuration, 
we set   
\begin{eqnarray}
&&\bvec{S}_1=\frac{d}{\sqrt{3}}(1, 1+\epsilon, -(1+\epsilon)),\\
&&\bvec{S}_2=\frac{d}{\sqrt{3}}(1, -1+\epsilon), 1+\epsilon),\\
&&\bvec{S}_3=\frac{d}{\sqrt{3}}(-1,-(1-\epsilon), -(1-\epsilon)),\\
&&\bvec{S}_4=\frac{d}{\sqrt{3}}(-1, (1-\epsilon), (1-\epsilon)),
\end{eqnarray}
where $\epsilon$ is taken to be an enough small value.

The VAP wave functions for the $0^+_1$, $3^-_1$, and  $4^+_1$ states 
have maximum overlaps with the $J^\pi$-projected $T_d$ wave function 
and that for the $1^-_1$ state has maximum overlap with 
the $J^\pi$-projected $T_d(F)$ wave function at a 
finite size $d=1.5-1.7$ fm.
Table \ref{tab:tetra} shows
calculated values of the $T_d$ and $T_d(F)$ components  in the 
VAP and VAP+$\alpha$GCM wave functions at $d=1.6$ fm.
It also shows the components in the single-base VAP wave function 
$\Phi^{J\pi}(\bvec{Z}^{\rm opt}_{J^\pi_1})$ without configuration mixing of the VAP 
wave functions.
The $0^+_1$ and $3^-_1$ states contain significant $T_d$ component
as 90\% and $60-70\%$, respectively, leading to an interpretation that they are approximately 
regarded as the $T_d$ band members. However, there is no $4^+$ state having 
dominant $T_d$ component.
Since the single-base VAP wave function for the $4^+_1$ state 
has a distorted tetrahedral intrinsic structure, two $4^+$ states obtained from $K=0$ and $K=2$
components by the $J^\pi$ projection share the $T_d$ component. These two $4^+$ states 
correspond to the  $4^+_1$ and $4^+_3$ states in the VAP result. 
In the VAP+$\alpha$GCM result, the $T_d$ component is fragmented further 
because of mixing with $^{12}\textrm{C}+\alpha$ components.
As a result of the distortion from the $T_d$ symmetry of the intrinsic $4\alpha$ 
structure and the mixing with $^{12}\textrm{C}+\alpha$ components
the $T_d$ component of the $4^+_2$ state
is reduced to 16\% in the VAP+$\alpha$GCM result.
For the vibration $T_d(F)$ mode, the obtained $1^-_1$ state contains significant 
$T_d(F)$ component as $50-60\%$ meaning that the $1^-_1$ state can be
roughly categorized into the $T_d(F)$ band.

Elliott assigned the $0^+_1$, $3^-_1$, and $4^+_2$ states as the $T_d$ band \cite{Elliott:1985dfe},
whereas Bijker and Iachello assigned the $0^+_1$, $3^-_1$, and $4^+_1$ states as the 
 $T_d$ band
and the $1^-_1$ state as the vibration  $T_d(F)$ band \cite{Bijker:2014tka,Bijker:2016bpb}.
For the $0^+_1$, $3^-_1$, 
and $1^-_1$ states, 
our result is approximately consistent with their assignment.
However for the $4^+$ state,  it is indicated that 
the $T_d$ symmetry is not stable at $J^\pi=4^+$ and its identity does not remain in $4^+$ states of $^{16}$O.
As discussed previously, we assigned the $0^+_1$, $3^-_1$, and $4^+_2$ states 
as the ground band members  
in the VAP+$\alpha$GCM  because they contain 
more than 55\% components of the corresponding VAP wave functions, which clearly 
show the similar tetrahedral $4\alpha$ structure. 
Our assignment of the $4^+_2$ to the ground band is consistent with that by Elliott, but 
the ground band is constructed 
from tetrahedral $4\alpha$ structure somewhat distorted from the ideal $T_d$ symmetry in the present result. 
It should be also noted again that 
significant state mixing occurs between $4^+_1$ and $4^+_2$ states.

\begin{table}[ht]
\caption{
\label{tab:tetra}
Tetrahedral $4\alpha$ component in the $^{16}$O wave functions
obtained by the VAP and VAP+$\alpha$GCM calculations. Calculated overlaps 
with the $T_d$ and $T_d(F)$ wave functions with the size $d=1.6$ fm are shown.
}
\begin{center}
\begin{tabular}{ccccccc}
\hline
&\multicolumn{2}{c} {1-base VAP} & \multicolumn{2}{c} {VAP} &\multicolumn{2}{c} {VAP+$\alpha$GCM} \\		
	&	$E_x$	&	$T_d$	&	$E_x$	&	$T_d$	&	$E_x$	&	$T_d$	\\	
$0^+_1$	&	0.0 	&	0.92 	&	0.0 	&	0.91 	&	0.0 	&	0.89 	\\	
$0^+_2$	&		&		&	13.1 	&	0.01 	&	9.7 	&	0.01 	\\	
$3^-_1$	&	9.4 	&	0.69 	&	8.3 	&	0.65 	&	7.6 	&	0.61 	\\	
$3^-_2$	&		&		&	18.3 	&	0.05 	&	13.4 	&	0.02 	\\	
$4^+_1$	&	17.1 	&	0.36 	&	16.1 	&	0.29 	&	13.7 	&	0.09 	\\	
$4^+_2$	&	21.9 	&	0.20 	&	18.2 	&	0.02 	&	14.5 	&	0.16 	\\	
$4^+_3$	&		&		&	21.0 	&	0.28 	&	16.6 	&	0.05 	\\	
$4^+_4$	&		&		&		&		&	17.7 	&	0.02 	\\	
$4^+_5$	&		&		&		&		&	18.0 	&	0.16 	\\	
	&	$E_x$	&	$T_d(F)$	&	$E_x$	&	$T_d(F)$	&	$E_x$	&	$T_d(F)$	\\	
$1^-_1$	&	11.0 	&	0.65 	&	10.3 	&	0.60 	&	9.4 	&	0.52 	\\	
$1^-_2$	&		&		&	17.0 	&	0.04 	&	12.1 	&	0.06 	\\	
\hline
\end{tabular}
\end{center}
\end{table}


\subsection{$^{12}\textrm{C}+\alpha$ cluster feature}

\begin{figure}[th]
\epsfxsize=0.3\textwidth
\centerline{\epsffile{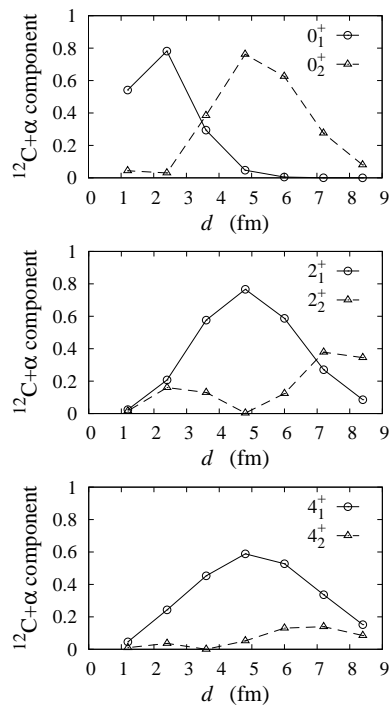}}
\caption{\label{fig:overlap1}
$^{12}$C($0^+_1)$+$\alpha$ component in the positive-parity states obtained by the VAP+$\alpha$GCM.
}
\end{figure}

\begin{figure}[th]
\epsfxsize=0.3\textwidth
\centerline{\epsffile{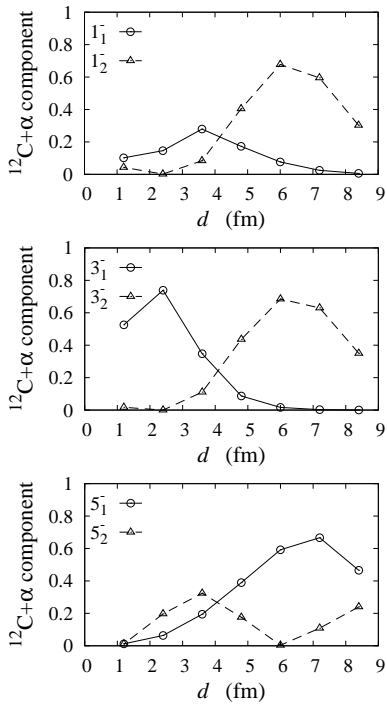}}
\caption{\label{fig:overlap2}
Same as Fig.~\ref{fig:overlap1} but for the negative-parity states.
}
\end{figure}

Figures~\ref{fig:overlap1} and \ref{fig:overlap2} show 
$^{12}\textrm{C}(0^+_1)+\alpha$ component in the 
positive- and negative-parity states, respectively. The component is
calculated by overlap with a $^{12}\textrm{C}(0^+_1)+\alpha$ cluster wave function at a certain distance
($d$) in the same way as Ref.~\cite{Yoshida:2016cfu}. 
The $^{12}\textrm{C}(0^+_1)$ cluster configuration is determined 
in an asymptotic region ($d=8.4$ fm is chosen in the present case) 
by diagonalization within a fixed-$d$ model space of
\begin{equation}
\Phi^{J\pi K}_{^{12}{\rm C}+\alpha}(d=8.4\textrm{ fm},\Omega_j,\bvec{Z}^\textrm{opt}_{\beta_\textrm{C}}).
\end{equation}
We truncate intrinsic configurations of $^{12}\textrm{C}$ as $\beta_\textrm{C}=0^+_1$ and $0^+_2$ 
for simplicity. 
In Figs.~\ref{fig:overlap1} and \ref{fig:overlap2}, 
the calculated result for positive- and negative-parity states
of the VAP+$\alpha$GCM is shown as functions of $d$. 
The $0^+_2$, $2^+_1$, and $4^+_1$ states have large
$^{12}\textrm{C}(0^+_1)+\alpha$ component with maximum amplitude
around $d=5$ fm, indicating that these states have the spatially developed 
$^{12}\textrm{C}(0^+_1)+\alpha$ cluster structure. 
The $1^-_2$, $3^-_2$, and $5^-_1$ states also 
have large $^{12}\textrm{C}(0^+_1)+\alpha$ component with maximum amplitude
around $d=6-7$ fm and show further development of the 
$^{12}\textrm{C}(0^+_1)+\alpha$ cluster structure. 
It indicates that these states belong to the $K^\pi=0^\pm$ bands constructed from the
$^{12}\textrm{C}(0^+_1)+\alpha$ structure,
consistently with the strong in-band $E2$ transitions.

In Table \ref{tab:a-decay}, calculated $\alpha$-decay widths are
compared with experimental data. 
We also show the theoretical values of a semi-microscopic 
$^{12}\textrm{C}+\alpha$ calculation in Ref.~\cite{Suzuki:1976zz2}.
The reduced widths are  
evaluated from the $^{12}\textrm{C}(0^+_1)+\alpha$ component at
a channel radius $a$ \cite{Kanada-Enyo:2014mri}. 
The squared dimensionless $\alpha$-decay width $\theta^2_\alpha$ evaluated from 
the experimental $\alpha$-decay widths are remarkably large  
for the $0^+_2$, $2^+_1$, $4^+_1$, $1^-_2$, $3^-_2$, 
$5^-_1$ states. The calculated $\theta^2_\alpha$ values at $a=6.0$ fm 
reasonably agree with the experimental values and also 
are consistent with the theoretical values of Ref.~\cite{Suzuki:1976zz2}.

\begin{table}[ht]
\caption{$\alpha$-decay properties. 
Experimental $\alpha$-decay energies ($E_\alpha$(MeV)), the widths $\Gamma$(MeV), 
and the squared dimensionless reduced widths $\theta^2_\alpha$. The experimental values of 
$\theta^2_\alpha$ are reduced by assuming the single channel $^{12}\textrm{C}(0^+)+\alpha$ decay. 
Theoretical values are calculated by the VAP+$\alpha$GCM.
The values of the $^{12}\textrm{C}+\alpha$ OCM calculation in Ref.~\cite{Suzuki:1976zz2} 
are also shown. The channel radii $a=4.8$ fm and $a=6$ fm are chosen.
\label{tab:a-decay}
}
\begin{center}
\begin{tabular}{ccccccccc}
		&$E_\alpha$ &		$\Gamma$ &	
 \multicolumn{4}{c} {$\theta_\alpha^2$} \\
&&& \multicolumn{2}{c}{exp.}&\multicolumn{2}{c}{cal.}& Ref.~\cite{Suzuki:1976zz2}\\	
&&&	$a=4.8$	&	$a=6$	&	$a=4.8$	&	$a=6$ &	$a=6$\\	
$0^+_2$	&	$-$1.11 	&		&		&		&	0.23 	&	0.19 &	\\
$0^+_3$ 	&	4.89 	&	1.5	&	0.00036	&	0.00039	&	0.005 	&	0.032 	& 0.075\\
$2^+_1	$	&	$-$0.24 	&		&		&		&	0.23 	&	0.18 & \\
$2^+_2	$	&	2.68 	&	0.62	&	0.0018	&	0.00090	&	0.001 	&	0.037 &0.0063 	\\
$2^+_3	$	&	4.36 	&	71	&	0.035	&	0.029	&	0.002 	&	0.003 	& 0.039\\
$4^+_1	$ 	&	3.19 	&	26	&	0.60	&	0.14	&	0.18 	&	0.16 &0.21	\\
$4^+_2$		&	0.28	&	0.0018	&	0.00054	&	0.016 	&	0.039 & 0.023	\\
$1^-_1$		&	$-$0.05 	&		&		&		&	0.052 	&	0.023 &	\\
$1^-_2$	 	&	2.42 	&	420	&	0.76	&	0.48	&	0.12 	&	0.20 & 0.33	\\
$3^-_1$		&	$-$.03 	&		&		&		&	0.026 	&	0.005 &	\\
$3^-_2$		&	4.4 	&	800	&	0.79	&	0.47	&	0.13 	&	0.21 &0.29	\\
$5^-_1$	&	7.50 	&	670	&	0.69	&	0.31	&	0.12 	&	0.18 &0.16	\\
\end{tabular}
\end{center}
\end{table}

\begin{figure}[th]
\epsfxsize=0.4\textwidth
\centerline{\epsffile{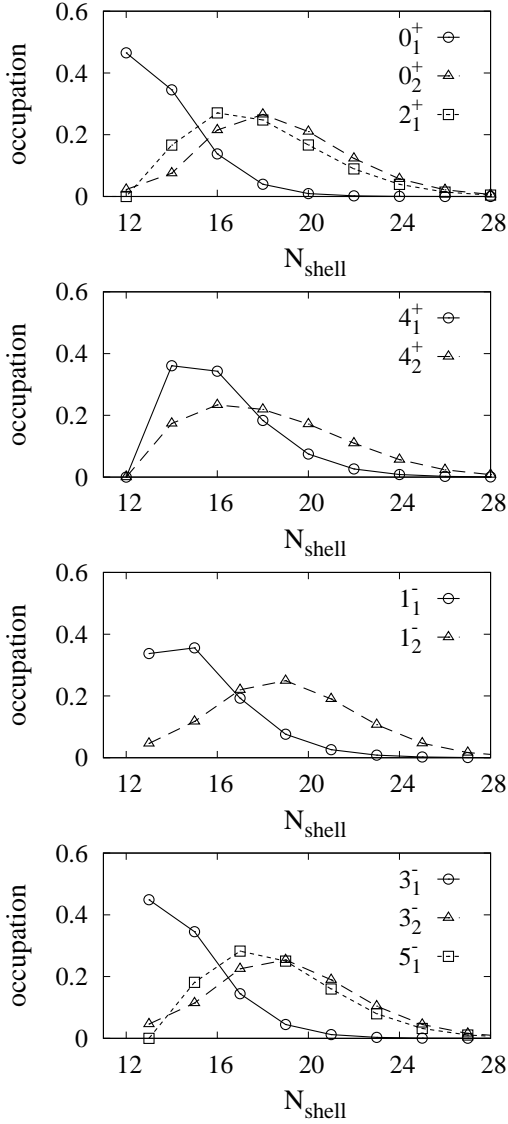}}
\caption{\label{fig:ho-vap}
Occupation probability of $N$ shells in the harmonic oscillator expansion 
in the states obtained by the VAP. 
}
\end{figure}

\begin{figure}[th]
\epsfxsize=0.4\textwidth
\centerline{\epsffile{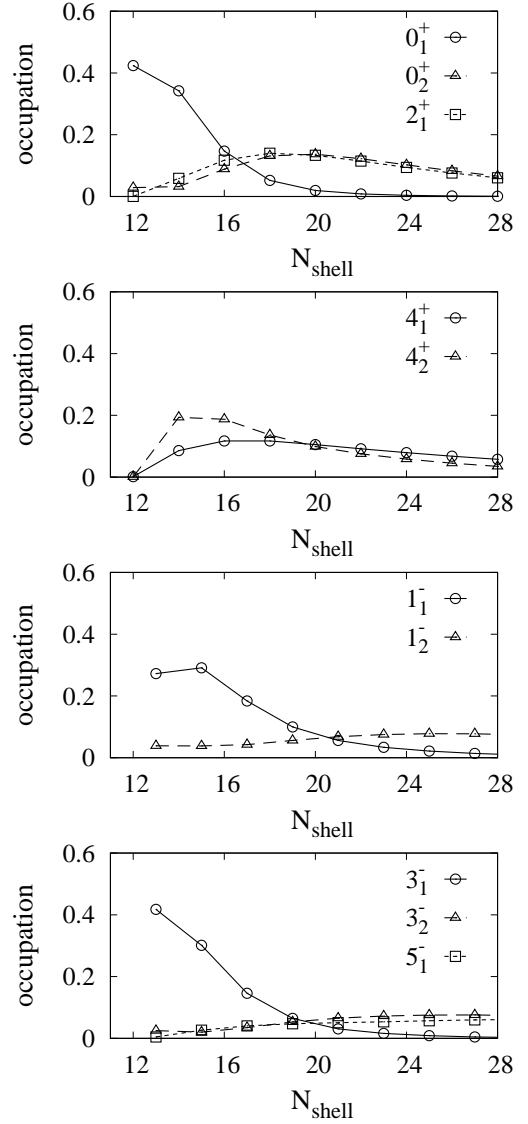}}
\caption{\label{fig:ho-gcm}
Occupation probability of $N$ shells in the harmonic oscillator expansion 
in the states obtained by the VAP+$\alpha$GCM. 
}
\end{figure}

Figures \ref{fig:ho-vap} and  \ref{fig:ho-gcm} show 
occupation probability of oscillator quanta $N$ shells
in a harmonic oscillator basis expansion. Here we use the size parameter 
$b=1/\sqrt{2\nu}$ of the harmonic oscillator.
For the $^{12}\textrm{C}+\alpha$ cluster states, the occupation probability is 
distributed widely in a higher shell region. In particular, the distribution is very 
broad in the VAP+$\alpha$GCM result because of the spatially developed  
$^{12}\textrm{C}+\alpha$ cluster structure. 
The $0^+_1$, $1^-_1$, $3^-_1$, and $4^+_1$ states of the VAP result
are dominated by the $N=12$, 13, 13, 14 shell component corresponding to the 
$0p0h$, $1p1h$,  $1p1h$,  and $2p2h$ on the $p$ shell, respectively. However, they also contain
more than 50\% higher shell components, which come from cluster correlations
in the finite size $4\alpha$ configurations.
The higher shell mixing in these states becomes large in the VAP+$\alpha$GCM result.
It should be stressed that the
significant higher shell mixing in the $0^+_1$ state, i.e.
the ground state cluster correlation enhances $B(E3;0^+_1\to 3^-_1)$
and $B(E3;0^+_1\to 4^-_1)$ as mentioned previously.

\section{Summary and outlooks}\label{sec:summary}
We have investigated structures of the ground and excited states of $^{16}\textrm{O}$
with the AMD+VAP method combined with the $^{12}\textrm{C(AMD)}+\alpha$ GCM.
The present result reasonably reproduces the experimental data of
energy spectra,  $E2$, $E3$, $E4$, and $IS1$ transitions as well as $\alpha$-decay properties.

The formation of 4 $\alpha$ clusters has been confirmed from nucleon degrees of freedom 
 in the present calculation 
without assuming existence of any clusters.
They form the (distorted) tetrahedral $4\alpha$ structure in the low-lying states, 
$0^+_1$, $3^-_1$, $4^+$, $1^-_1$, and $2^-_1$, and the
$^{12}\textrm{C}+\alpha$ cluster structures in the excited states near and above the 
$^{12}\textrm{C}+\alpha$ threshold.

The $0^+_1$, $3^-_1$, and $4^+_2$ states are assigned to the ground band constructed from 
the tetrahedral 4$\alpha$ structure.
The tetrahedral 4$\alpha$ structure does not necessarily have the ideal tetrahedral configuration 
with the $T_d$ symmetry, but a somewhat distorted tetrahedral shape. 
Nevertheless, 
the $0^+_1$ and $3^-_1$ have significant (90\% and 60\%) component of the 
$T_d$-invariant $4\alpha$ configuration projected onto the $J^\pi$ eigen state, 
and therefore, they can be 
approximately interpreted as the $T_d$ band members. 
In $4^+$ states, 
the $T_d$ component is shared mainly by two $4^+$ states because of the distortion of the
tetrahedral shape from the $T_d$ symmetry, and fragmented further by mixing of $^{12}\textrm{C}+\alpha$ states.
It indicates that the tetrahedral $4\alpha$ structure may be rather fragile at $J^\pi=4^+$, 
and the ideal $T_d$ member with $4^+$ does not appear in $^{16}$O.
The $1^-_1$ state can be roughly categorized into the vibration mode, $T_d(F)$ band.

Our assignment of the $4^+_2$ to the ground band is consistent with that of Ref.~\cite{Elliott:1985dfe}.
However, the $4^+_2$ state is not the ideal $T_d$ member but 
has the distorted tetrahedral $4\alpha$ shape as the dominant component. It should be also noted that 
significant state mixing occurs between $4^+_1$ and $4^+_2$ states.

The $^{12}\textrm{C}+\alpha$ cluster structure constructs
the $K^\pi=0^+$ band consisting of the $0^+_2$, $2^+_1$, $4^+_1$
and the $K^\pi=0^-$ band of $1^-_1$, $3^-_2$, $5^-_1$. These states contain
the dominant $^{12}\textrm{C}(0^+_1)+\alpha$ component and large 
$\alpha$-decay widths, which are consistent with experimental observations.
The present result for the  $K^\pi=0^\pm$ $^{12}\textrm{C}(0^+_1)+\alpha$ bands are consistent with those of 
the semi-microscopic and microscopic cluster model calculations 
\cite{Suzuki:1976zz,Suzuki:1976zz2,Descouvemont:1987uvu,Descouvemont:1991zz,Descouvemont:1993zza}.
The present assignment of the $4^+_1$ state to the $^{12}\textrm{C}+\alpha$ band is 
supported by experimental data of the strong $E2$ transition to the $2^+_1$ and the large $\alpha$-decay width.

The $E3$ and $E4$ transition strengths have been discussed.
The calculated $B(E3;3^-\to 0^+_1)$ is in good agreement with the experimental data.
The $E3$ strength is enhanced because of the tetrahedral $4\alpha$ structure in the $0^+_1$ and $3^-_1$ states. 
The ground state $4\alpha$ correlation plays an important role in the enhancement of the $E3$ strength.
For the $E4$ strength, the present calculation reproduces well the 
experimental  $B(E4;4^+_1 \to 0^+_1)$. Historically, the significant $B(E4)$ measured by $(e,e')$ scattering has often drown 
attention to cluster structure of the $4^+_1$, which could be the ground band member with the $T_d$ symmetry.  
In the present result of the VAP+$\alpha$GCM, the $4^+_1$ belongs to not the ground band but 
the $^{12}\textrm{C}+\alpha$ band starting from the $0^+_2$ state. Although, inter-band transitions are  
generally weak, however, the $B(E4;4^+_1 \to 0^+_1)$ is increased by 
the significant state mixing of the 
$^{12}\textrm{C}+\alpha$ and tetrahedral 4$\alpha$ structures between $4^+_1$ and
$4^+_2$ states and also by slight mixing of the $^{12}\textrm{C}+\alpha$ component in the $0^+_1$.
As a result, the calculated $B(E4;4^+_1 \to 0^+_1)$ is as large as the experimental data
in spite of the different structures in the initial and final states.
In the experimental measurement using $(e,e')$ scattering \cite{Buti:1986zz}, it has been reported that 
the $E4$ transition strength for $4^+_2 \to 0^+_1$ is the same order as that for $4^+_1 \to 0^+_1$. 
It may support significant state mixing between two $4^+$ states.

In the traditional microscopic cluster model calculations with the Volkov interaction (a density-independent two-body 
interaction), it has been known that 
excitation energies of the $^{12}\textrm{C}+\alpha$ cluster states are highly overestimated.
The excitation energies of $^{12}\textrm{C}+\alpha$ cluster states are largely improved in the present 
calculation. One of the main reasons is that we used the effective interaction with the zero-range three-body term, with which 
the $\alpha$-decay threshold energy is reproduced. 
Internal excitation and angular momentum projection of the subsystem $^{12}$C cluster 
also give significant contribution to the energy reduction of  the $^{12}\textrm{C}+\alpha$ cluster states.
However, the theoretical excitation energies is still higher than the experimental data.
For better reproduction of the experimental energy spectra, 
further improvement of the model space with more sophisticated effective interactions including  
the tensor force may be necessary.
Moreover, the present calculation is based on the bound state approximation.
Coupling with continuum states should be carefully taken into account to discuss 
detailed properties of resonances. In particular, since the state mixing 
between the $4^+_1$ and $4^+_2$ is very sensitive to their relative energy positions, 
further improvement and fine tuning of the model calculation are needed to 
discuss detailed properties of these states.

\section*{Acknowledgments}
The author thanks to Dr. Y. Hidaka for fruitful discussions.
The computational calculations of this work were performed by using the
supercomputers at YITP.
This work was supported by Grant-in-Aid for Scientific Research from Japan Society for the Promotion of Science (JSPS) 
Grant Number 26400270.
It was also supported by 
the Grant-in-Aid for the Global COE Program ``The Next Generation of Physics, 
Spun from Universality and Emergence'' from the Ministry of Education, Culture, Sports, Science and Technology (MEXT) of Japan.

\end{document}